# TomoSphero: Fast Differentiable Projector for Planetary and Solar Tomography on Spherical Grids

Evan Widloski
evan_apj@widloski.com

Lara Waldrop
lwaldrop@illinois.edu

*Electrical and Computer Engineering*
*University of Illinois Urbana-Champaign*

***Abstract*— Computational tomography is a tool for determining the internal structure of objects from a set of projections, typically taken along some regular path. In recent years, methods and GPU-accelerated libraries have emerged that allow for fast reconstruction from projections along more complicated paths. Most of these libraries rely on a Cartesian discretization of the object, which is not appropriate for all scenarios. We present TomoSphero, a differentiable tomographic projector over spherical grids which are often used in planetary and solar tomography. TomoSphero is designed to be used as a building block in reconstruction algorithms and includes common projection types such as cone-beam and parallel-beam, but is flexible enough to accommodate arbitrary projections. TomoSphero is implemented in PyTorch which allows for fast projection computation on GPUs, easy access to modern machine learning optimizers, and automatic differentiation for rapid prototyping of parametric models.**

## 1. Introduction

Tomography is a method for determining the internal structure of objects from a set of measurements that penetrate into the object being measured. These measurements (sometimes called projections or sinograms) are usually captured from a variety of locations and times which are collectively referred to as the *view geometry*. Measurements are typically modeled as

$$\boldsymbol{y} = F\boldsymbol{\rho} + \boldsymbol{\varepsilon} \tag{1}$$

where $\boldsymbol{y}$ is a collection of measurements, $F$ is a linear projection operator, $\boldsymbol{\rho}$ is the object under study, and $\boldsymbol{\varepsilon}$ is noise.

Tomography has found application in a vast number of domains such as medical imaging, crystallography, and remote sensing, especially coronal electron density [1], [2], [3], [4], [5]. In this paper we present TomoSphero[1], a Python library for planetary and solar tomography.

Fast tomographic reconstruction algorithms that implement explicit inversion formulas typically work only for specific view geometries (such as circular or helical view geometry) and are referred to as *filtered back projection* (FBP) algorithms [6]. However, some situations necessitate more complicated measurement schemes than are allowed by FBP-type algorithms. Some examples include quality control on a factory assembly line, where X-ray measurements of large components are taken from arbitrary positions by a robotic arm [7], or spaceborne sensors whose view geometries are determined by orbital parameters of a spacecraft [8], [9]. For these situations requiring more flexible view geometries where an exact inverse solution is not available, *iterative reconstruction* (IR) algorithms prevail, usually solving an optimization problem of the form

$$\hat{\boldsymbol{\rho}} = \arg\min_{\boldsymbol{\rho}} \|\boldsymbol{y} - F\boldsymbol{\rho}\|_2^2 + ... \tag{2}$$

Examples include SIRT [10], TV-MIN [11], ART [12], CGLS [13], Plug-and-play [14] and many others. These algorithms obtain synthetic projections of a candidate object using an operator (sometimes called a *raytracer*) that simulates waves traveling through the object medium. They produce a reconstruction by repeatedly tweaking the candidate object to minimize discrepancy between synthetic and actual projections, and they stand to benefit the most from a fast operator implementation. Most tomographic operators treat each pixel on the detector as an independent computational task, which has led to the development of tomography libraries that are capable of simultaneously utilizing multiple cores on a CPU or hardware accelerator.

In cases where a simultaneous computation for every pixel of every measurement would consume more memory than is available, some algorithms operate *out-of-core*, where they parallelize as many tasks as will fit into available memory, then serially queue the remaining tasks for processing after current tasks are complete. TomoSphero is not capable of out-of-core operation, so a characterization of its memory usage is also presented in Section 5.2.

Another consideration in tomographic reconstruction is the choice of grid type for discretization of the reconstructed object.

[1] https://github.com/Evidlo/tomosphero

Most publications consider a regular rectilinear grid, which is a reasonable choice when the underlying structure of the object is completely unknown or the scale of features is uniform throughout the object. However, in cases where some prior information is known about the location of high-detail regions within the object, or when symmetries in the view geometry exist [15], a good choice of coordinate system can sample the object more efficiently for decreased computational requirements and lower quantization errors. The primary focus of TomoSphero is in the domain of atmospheric tomography, where regular spherical grids are well-suited for modeling solar and planetary atmospheres that exhibit spherical symmetries [1], [2], [3].

Many reconstruction algorithms rely on gradient-based optimization to solve for an object whose structure corresponds to measurement data. Manually-coded gradients give an exact solution but can be time-consuming and error-prone to implement correctly, especially in a language designed for a hardware accelerator or if the problem involves a complex parametric object model (e.g. splines, wavelets, spherical harmonics). Finite differentiation and symbolic differentiation exist to solve for gradients for any generic problem, but can be extremely slow for high dimensional problems or can yield complicated expressions for the symbolic derivative (a.k.a. expression swell). Automatic differentiation (*autograd*) is a class of techniques that convert an arbitrary expression into a computational graph of simpler functions, then compute the overall derivative by applying the chain rule at each node. Modern machine learning libraries such as PyTorch [16] and Jax [17] provide such capabilities for building this computational graph. TomoSphero is implemented on top of PyTorch and its autograd capabilities enable rapid prototyping of different parametric models and regularizations with minimal code changes.

In this paper we describe TomoSphero, a tomographic projector implemented in Python designed for reconstruction problems on spherical grids. TomoSphero is automatically-differentiable, works on GPUs, and is easily integrable into machine learning methods through PyTorch.

In Section 4, we demonstrate its use on an example tomography problem and highlight visualization capabilities for setup validation and presentation.

TomoSphero development was motivated by the Carruthers Geocorona Observatory, a spacecraft containing UV imagers which will survey the Earth's exosphere. An example of retrieving a synthetic exosphere is provided in Section 6.

A comparison of TomoSphero's capabilities against other popular libraries is shown in Table 1.

| Name | Grid Type | GPU Support | Autograd | Visualization | Out-of-core |
|---|---|---|---|---|---|
| TIGRE [18] | Cartesian | Yes | No | No | Yes |
| LEAP [19] | Cartesian, analytic | Yes | Yes | No | Yes |
| ASTRA [20], [21], [22] | Cartesian | Yes | Yes | No | Yes |
| MBIRJAX [23] | Cartesian | Yes | Yes | No | No |
| ToMoBAR [24] | Cartesian | Yes | No | No | Yes |
| CIL [25] | Cartesian | Yes | No | Yes | Yes |
| Tomosipo [26] | Cartesian | Yes | Yes | Yes | Yes |
| TomoSphero | Spherical | Yes | Yes | Yes | No |

Table 1: Non-exhaustive overview of other tomography libraries

# 2. Tomography Problem

In this section we establish mathematical notation used in the rest of the paper, provide a formulation and describe assumptions of the emission model, and define the spherical discretization grid. A summary of terms and symbols used in this section can be found in Section 9.1.

## 2.1. Emission Model

Most tomography applications assume that measurements may be modeled as a simple integral along a line through the object known as a *line of sight* (LOS) or a *ray*. In the context of atmospheric optical tomography, illuminated gas along the LOS scatters (or absorbs and reemits) light in the direction of the sensor where it is detected. We assume the photons are only scattered once after being emitted by the source of light (e.g. the Sun) and before entering the detector, an approximation that is valid in the *optically thin* regime of the atmosphere [27].

This integral is known as a Fredholm integral of the first kind and is given below for a single LOS

$$y = F\boldsymbol{\rho} = \int_{l=0}^{\infty} \boldsymbol{\rho}\left(\vec{x} + \vec{n}l\right) \mathrm{d}l \tag{3}$$

where $\boldsymbol{\rho}$ represents the 3D object, $\vec{x}$ is the ray starting point (the detector), $\vec{n}$ is the ray direction and $l$ is distance along the ray from the starting point.

To compute the integral in Equation 3 numerically, we must discretize the object over a spherical grid and rewrite the integral as a sum, which can be compactly described as an inner product

$$y = \sum_{i=0}^{N-1} \rho_i \cdot \Delta l_i = \langle \rho, \Delta l \rangle \tag{4}$$

where $N$ is the number of spherical grid cells (a.k.a. *voxels*) intersecting the ray, $\rho_i$ is the object value within each cell, and $\Delta l_i$ is the intersection length as illustrated in Figure 1. The arrays $\rho$ and $\Delta l$ have length $N$ and are described in more detail in Section 3.

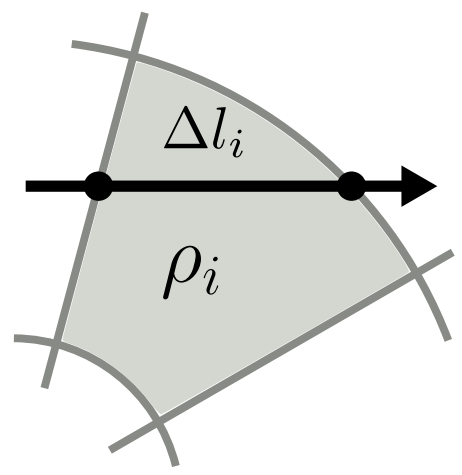

Figure 1: 2D representation of a ray intersecting a single spherical grid cell

### 2.2. Grid Definition

The grid determines the location of voxels in 3D space. In spherical coordinates the grid lines, or *grid boundaries* are spheres, cones, or planes that correspond to the radial, elevational, and azimuthal dimensions as in Figure 2. We refer to the space between two adjacent grid boundaries as a *grid region*. For a spherical grid of shape ($N_r$, $N_e$, $N_a$), there should be $N_r + 1$, $N_e + 1$, and $N_a + 1$ boundaries of each type, respectively.

The user-defined locations of the grid boundaries need not be uniform nor must they cover the extent of the entire sphere. This can be useful when spatially varying resolution is needed or only a small wedge needs to be raytraced (e.g. a local region of planetary atmosphere), as in Figure 3. Figure 4 shows a 2D representation of a grid of shape (1, 2, 7) that does not cover the entire sphere and will be used as an example throughout the next section.

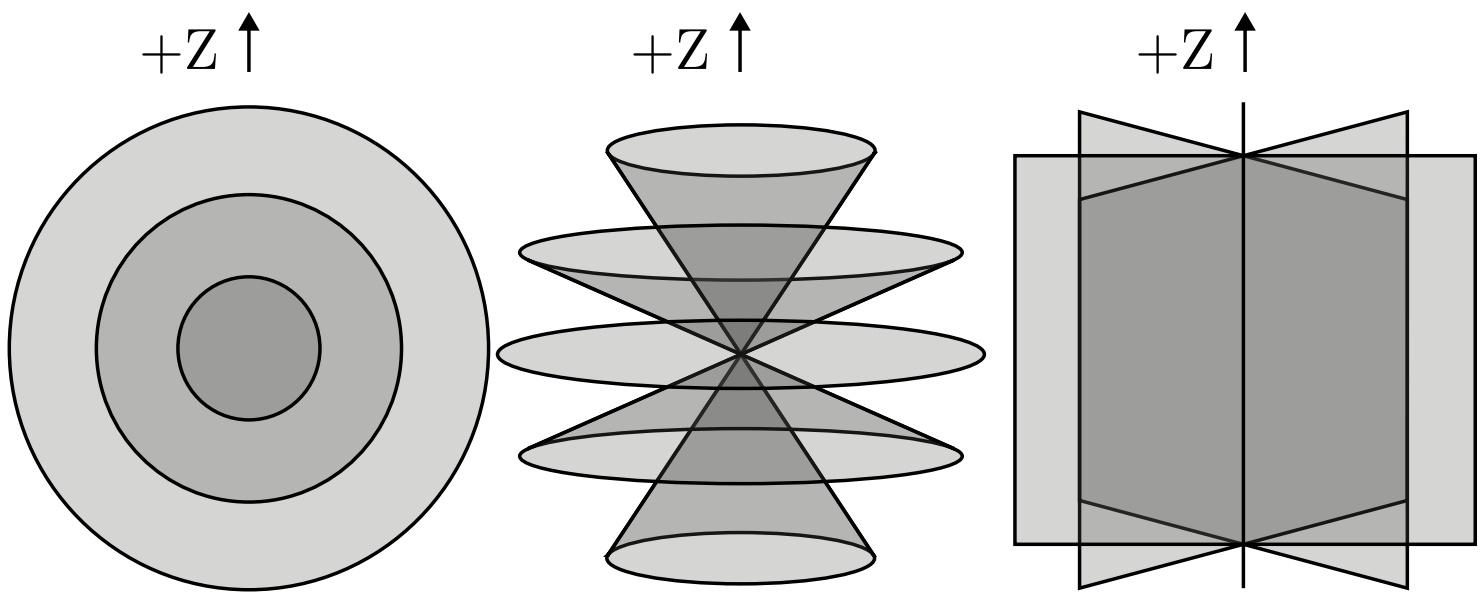

Figure 2: Radial, elevational, azimuthal grid boundaries for a grid with shape (2, 4, 8)

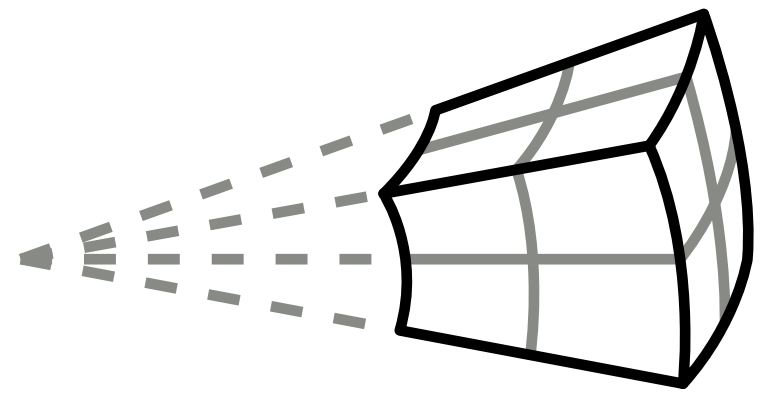

Figure 3: A spherical grid with shape (2, 2, 2) defined over a small wedge

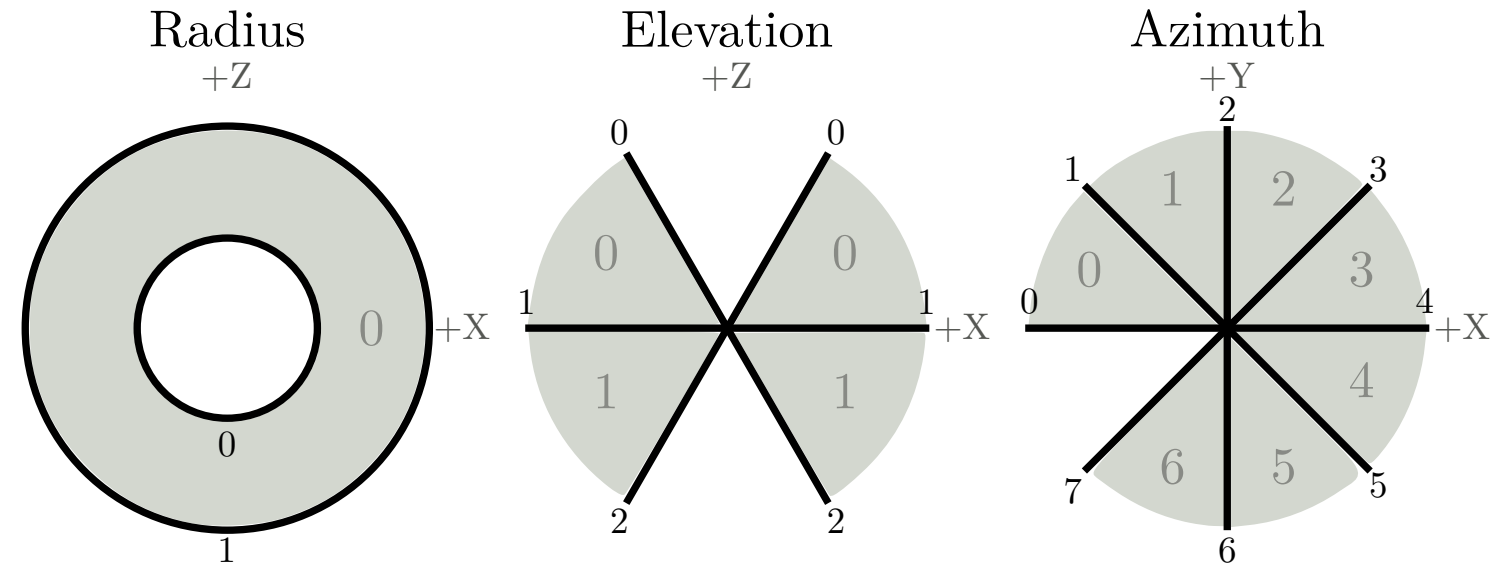

Figure 4: Sample grid of shape (1, 2, 7) broken into radial, elevational, and azimuthal boundaries and shown in 2D. An index is assigned to each boundary (in black) and each region (in gray). The shaded area is within the grid and white area outside.

# 3. Algorithm Outline

The steps for this raytracer can be broken into 4 parts, which are described in detail in this section:

1. For a given ray, compute intersection points with all boundaries and their distances from ray starting point
2. Determine crossing direction for all intersection points and compute region index
3. Sort intersection points and regions by distance from ray starting point and find lengths between points
4. Raytrace ray by computing inner product of ray lengths with values from voxels along ray

Since the focus of this paper is on iterative tomographic reconstruction where the view geometries are fixed and known, steps 1-3 can be precomputed and stored before reconstruction begins. This saves computation time at the expense of increased memory usage.

### 3.1. Step 1 - Ray-Boundary Intersection Points

The first step is to compute the intersection points of all rays with all boundaries and their distances from ray starting points. An array of intersection point coordinates and an array of intersection point distances is computed for each type of boundary separately. We consider only a single ray in this section without loss of generality.

It is desirable for the arrays allocated for ray-boundary intersection point coordinates and distances to be equal size for all rays to facilitate array programming and parallelization. A ray may intersect a sphere or a cone 2 times and a plane 1 time. Therefore, we assume an upper bound where each ray has exactly $2(N_r + 1)$, $2(N_e + 1)$, and $N_a + 1$ intersections with each type of boundary.

For the cases where a boundary has no intersection points with a ray, coordinates are still computed and stored for these points, but the associated value in the distance array $l$ is set to *inf*, as illustrated in Figure 5.

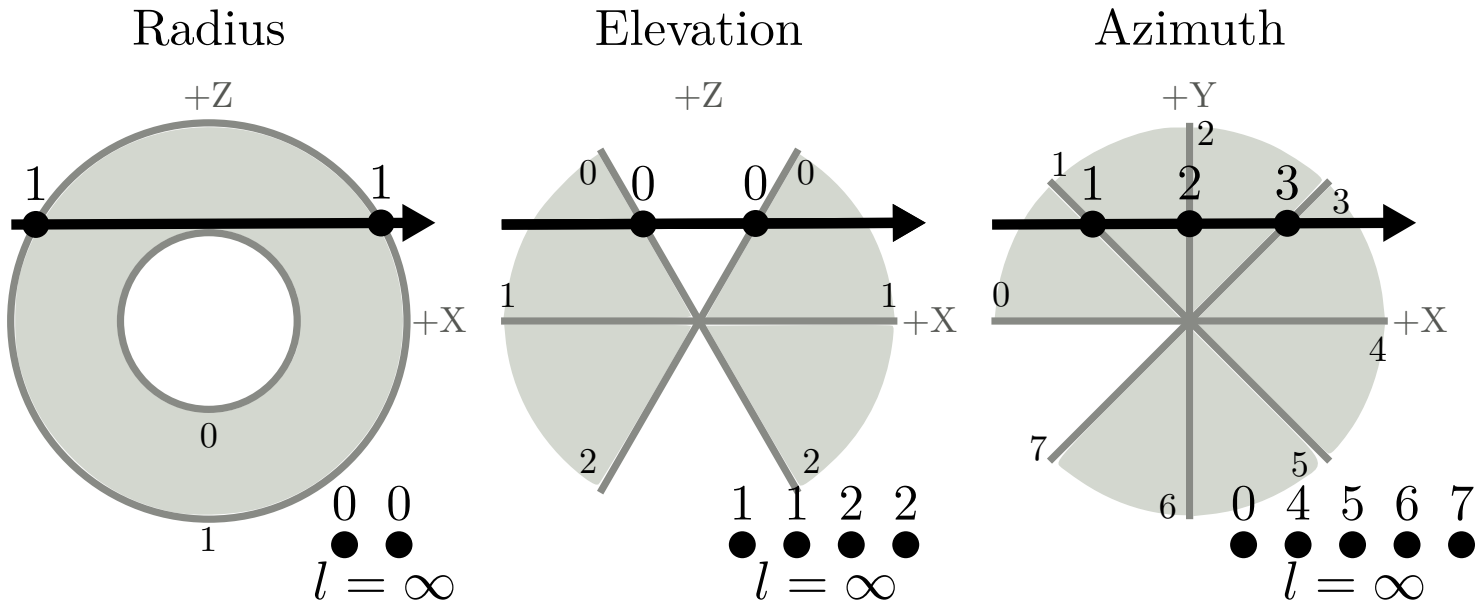

Figure 5: Intersection points along a LOS labelled with index of associated boundary. Points along bottom of figure are associated with non-intersecting boundaries.

The expressions for determining intersection points are given in [28], [29], [30].

### 3.2. Step 2 - Boundary Crossing Direction

Since the overall goal is to determine the indices of particular voxels the ray is crossing as it passes through the grid, we must convert each boundary index computed in the previous section to a region index. To do this, we must determine for each intersection whether the ray crosses the boundary in a "positive" direction (increasing boundary index) or a "negative" direction (decreasing boundary index) as shown in Figure 6. The algorithm for determining crossing direction is specific to each boundary type and depends only on the crossing point coordinate and ray direction. It is described in Section 9.2 in the appendix. Finally, boundary indices are converted to region indices by decrementing indices of crossing points with a negative direction, as shown in Figure 7 and Figure 8.

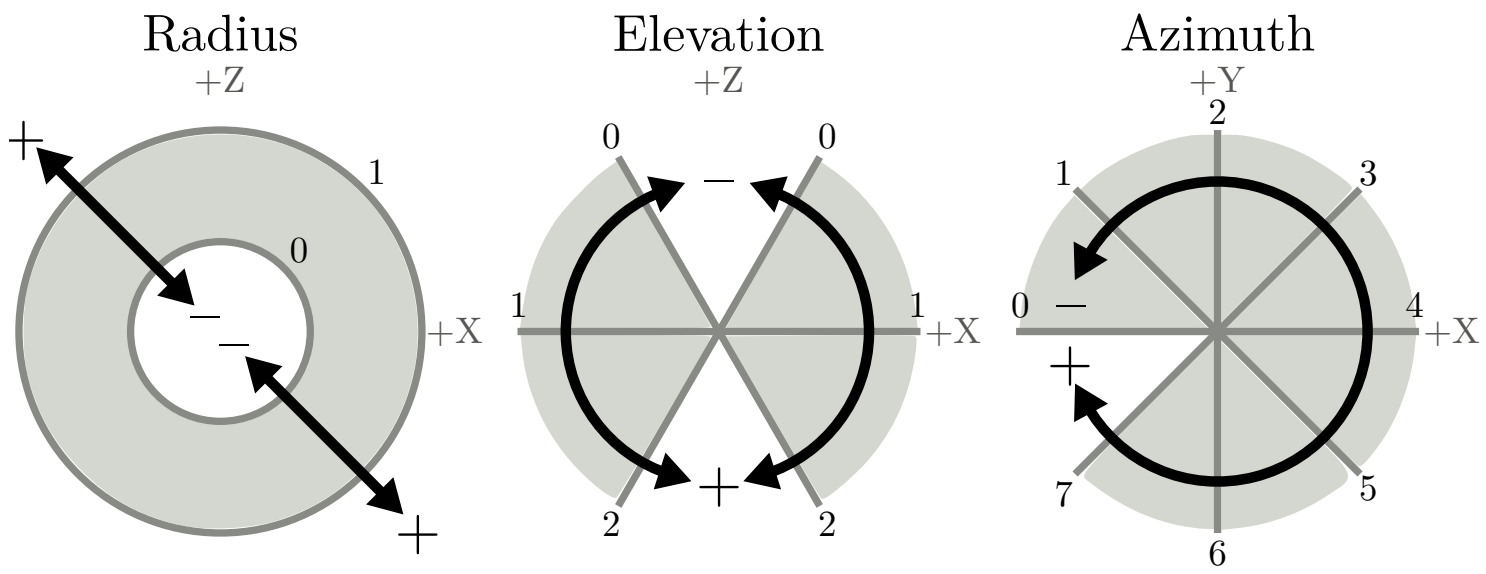

Figure 6: Positive and negative boundary crossing directions

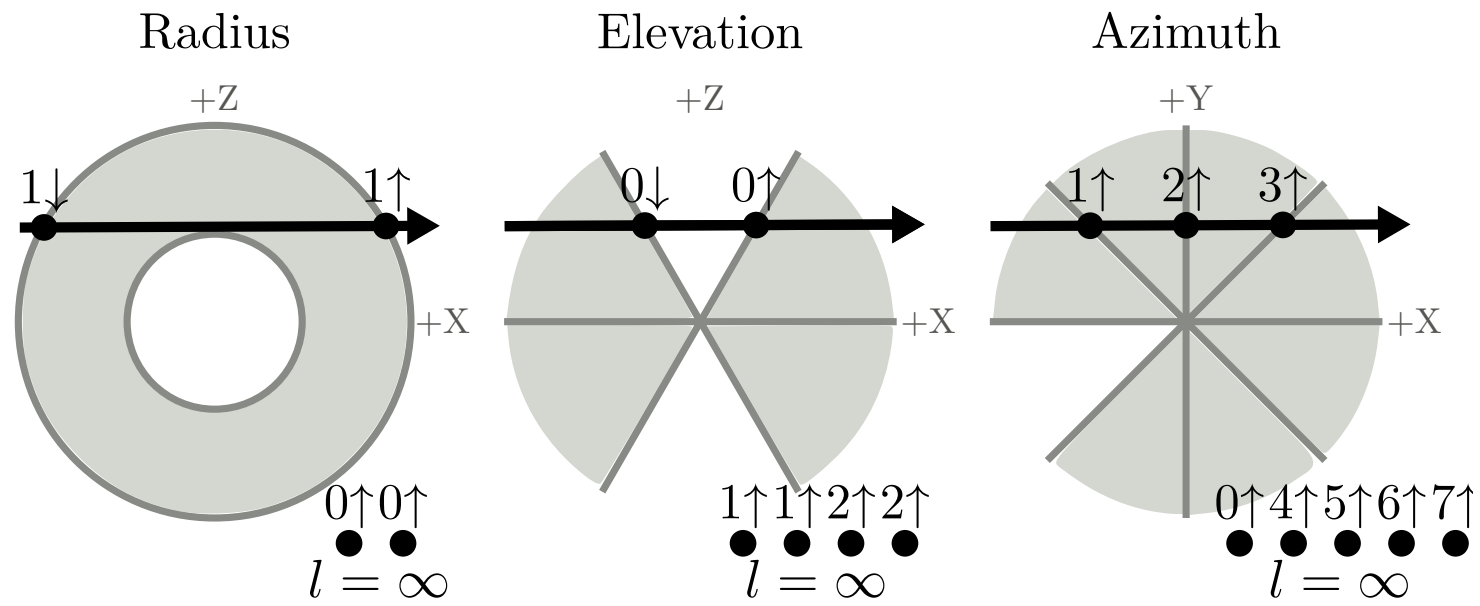

Figure 7: Intersection points along a LOS annotated with crossing direction (↑:positive, ↓:negative)

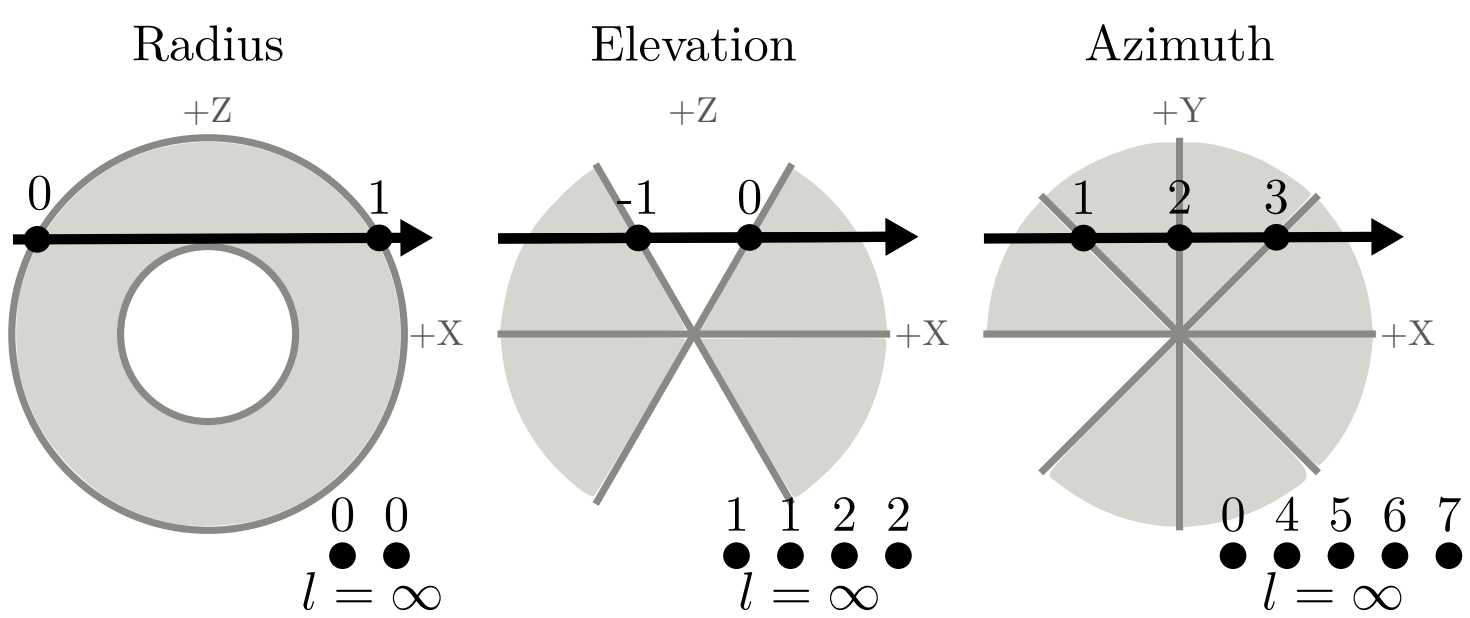

Figure 8: Region indices computed from crossing direction

### 3.3. Step 3 - Sorting Intersection Points

In this step we collect the region indices for all intersection points into a single list and sort the points by their distance $l$ from the ray start location, as illustrated in Figure 9. Next we compute the difference between adjacent points $\Delta l_i$ which represents the intersection length of the ray with a particular voxel. Note that this results in $\Delta l_i$ equal to NaN or $\infty$ for intersection points with $l = \infty$.

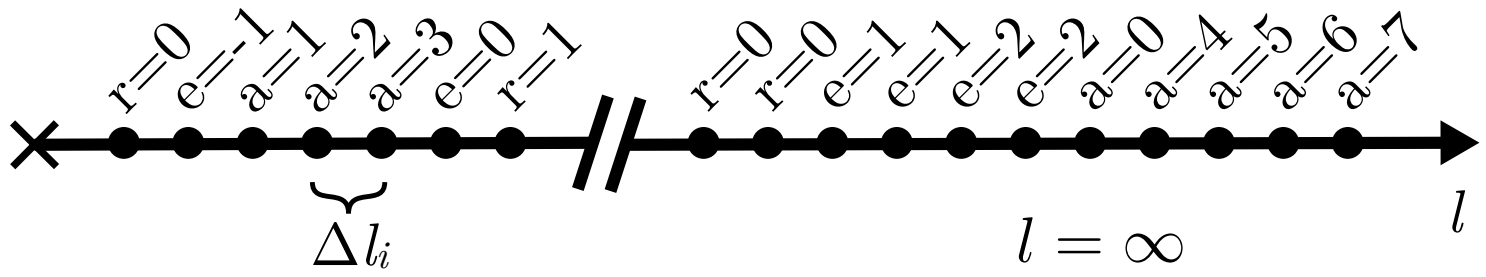

Figure 9: Intersection points along a LOS for all boundary types collected into a list and sorted by distance $l$ from ray start position, marked with x

Next, we convert our list of region indices into a list of voxel indices by computing the voxel index of the ray starting location. For each boundary crossed by the ray, the index of the next voxel changes in only a single dimension.

This step is illustrated in Table 2 (a), where the sorted region indices are inserted into an empty table starting with the ray starting point, then in Table 2 (b) missing indices are filled in from the previous row to form complete 3D voxel indices, for a total of $N = 2(N_r + 1) + 2(N_e + 1) + (N_a + 1)$ rows.

We also account here for invalid region indices generated in step 2 where the ray passes outside the area defined by the grid or when boundaries have no ray intersection. We achieve this by setting $\Delta l_i = 0$ for any row $i$ where the region index is invalid (less than 0, or greater than or equal to the corresponding $N_r$, $N_e$, $N_a$), or where $\Delta l_i$ is $\infty$ or NaN. Setting $\Delta l_i = 0$ negates any contribution of this row to the raytracing step. This is shown in Table 2 (c).

(a) Sorted

| r | e | a | $\Delta l$ |
|---|---|---|---|
| 1 | 0 | 0 | .3 |
| 0 | | | .8 |
| | −1 | | .7 |
| | | 1 | .5 |
| | | 2 | .5 |
| | | 3 | .2 |
| | 0 | | .7 |
| 1 | | | ∞ |
| 0 | | | NaN |
| 0 | | | NaN |
| | ... | | |
| | | 7 | NaN |

(b) Forward-filled

| r | e | a | $\Delta l$ |
|---|---|---|---|
| 1 | 0 | 0 | .3 |
| 0 | 0 | 0 | .8 |
| 0 | −1 | 0 | .7 |
| 0 | −1 | 1 | .5 |
| 0 | −1 | 2 | .5 |
| 0 | −1 | 3 | .2 |
| 0 | 0 | 3 | .7 |
| 1 | 0 | 3 | ∞ |
| 0 | 0 | 3 | NaN |
| 0 | 0 | 3 | NaN |
| | ... | | |
| 0 | 0 | 7 | NaN |

(c) Zeroed

| r | e | a | $\Delta l$ |
|---|---|---|---|
| 1 | 0 | 0 | 0 |
| 0 | 0 | 0 | .8 |
| 0 | −1 | | 0 |
| 0 | −1 | 1 | 0 |
| 0 | −1 | 2 | 0 |
| 0 | −1 | 3 | 0 |
| 0 | 0 | 3 | .7 |
| 1 | 0 | 3 | 0 |
| 0 | 0 | 3 | 0 |
| 0 | 0 | 3 | 0 |
| | ... | | |
| 0 | 0 | 7 | 0 |

Table 2: a) Sorted region indices are inserted into empty table along with voxel intersection lengths. b) Missing table entries are filled in from the previous row to form a complete 3D voxel index. c) Rows with invalid entries are set to zero so the row has no effect during raytracing.

### 3.4. Step 4 - Line Integration

The final step of the raytracer is to perform a weighted sum of the object values at the voxel indices given in the table in the previous section with the $\Delta l$ column as weights, compactly written as an inner product

$$y = F\boldsymbol{\rho} = \sum_{i=0}^{N-1} \boldsymbol{\rho}[r_i, e_i, a_i] \cdot \Delta l_i = \langle \boldsymbol{\rho}[r, e, a], \Delta l \rangle \tag{5}$$

where $\boldsymbol{\rho}$ is the object to be integrated and $r_i$, $e_i$, and $a_i$ are indices from the $i$-th row of the table.

## 4. API Overview

This library is designed to allow the user to easily construct tomographic projectors in just a few lines by instantiating classes that represent the grid, view geometries and forward raytrace operator. These classes also support visualization via the `.plot()` method to aid in validation and presentation. In this section, we provide some examples of setting up these classes for typical tomography problems as well as visualizations generated from the library.

### 4.1. Grid

The grid, which defines the physical extent and shape of the object to be traced, is created using the `SphericalGrid` class and may be defined by providing `shape` and `size` arguments when a uniform spherical grid is desired, as in Listing 1. Alternatively, one may manually specify the grid boundary locations via `r_b`, `e_b`, `a_b` arguments for a non-uniform grid.

```
grid = SphericalGrid(
  shape=(30, 30, 30), # voxels
  size_r=(0, 10)
)
grid.plot()
```

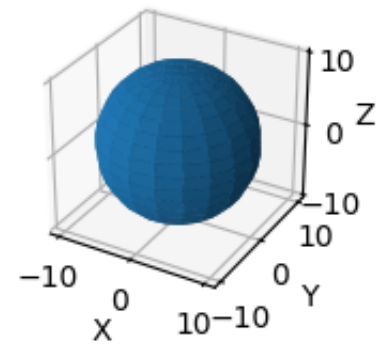

Listing 1: An origin-centered grid defined out to radius 10

### 4.2. View Geometries

A *view geometry* is a logical collection of lines of sight. This library provides a few built-in view geometries for common tomography paradigms shown in Listing 2 and Listing 3. In these examples, `pos` and `shape` control the detector position and shape. Detector orientation is inferred automatically, but this behavior may be overriden.

```
geom = ConeRectGeom(
  pos=(10, 0, 0),
  shape=(64, 64) # pixels
)
geom.plot()
```

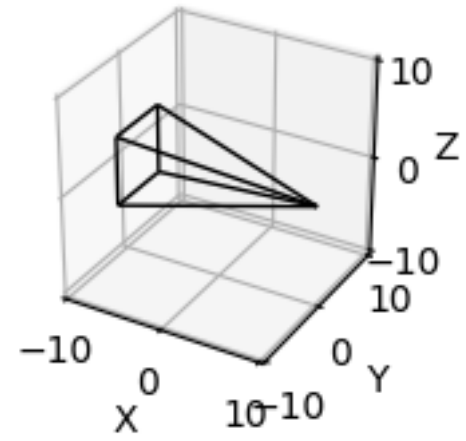

Listing 2: Cone beam view geometry with rectangular detector located at vertex

```
geom = ParallelGeom(
  pos=(10, 0, 0),
  shape=(4, 4) # pixels
)
geom.plot()
```

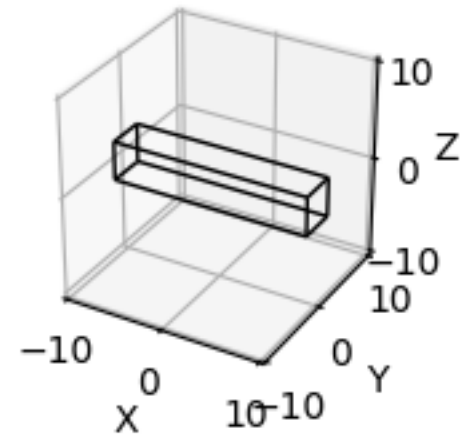

Listing 3: Parallel beam view geometry with rectangular detector

For more exotic detector geometries, one may define an entirely custom view geometry by specifying each ray manually via the `ViewGeom` class as in Listing 4 or by subclassing `ViewGeom`.

```
import torch as t
rays = t.rand((4, 4, 3)) / 5
rays[:, :, 0] = -1
ray_starts = t.tensor((10, 0,
0)).broadcast_to(rays.shape)
geom = ViewGeom(
  rays=rays,
  ray_starts=ray_starts
)
geom.plot()
```

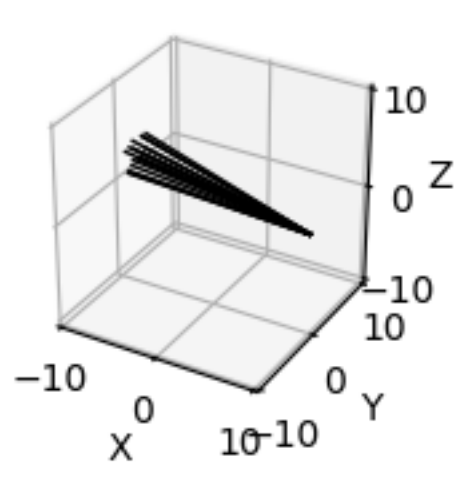

Listing 4: Custom view geometry for a 4-by-4 detector with random rays

### 4.3. Composing View Geometries

The view geometries given above can be composed into a `ViewGeomCollection` to create more complex geometries. A `ViewGeomCollection` may represent a set of sensors that are taking measurements at the same instant or a single moving sensor making sequential measurements. Programmatically, two view geometries can be combined by Python's addition operator, as in Listing 5. In this case, `.plot()` returns an animated visualization showing the sequence of view geometries.

```
geom = None
# sweep 360° around origin
for w in t.linspace(0, 2*t.pi, 50):
  pos=(5*t.cos(w), 5*t.sin(w), 1)
  # combine geoms by addition
  geom += ConeRectGeom(
    pos=pos,
    shape=(100, 100), # pixels
    fov=(25, 25) # degrees
  )
anim = geom.plot()
```

Listing 5: Circular orbit constructed from individual view geometries and a single frame from the resulting animation

### 4.4. Operator

When both view geometry and grid have been defined, the user may instantiate an `Operator` class as in Listing 6, which carries out the precomputation steps described in Section 3 and stores results in memory. When called with a 3D (static) or 4D (dynamic) object array, this instantiated `Operator` returns raytraced measurements. It is automatically-differentiable on account of PyTorch's autograd capability, which is useful in reconstruction algorithms that require access to gradients.

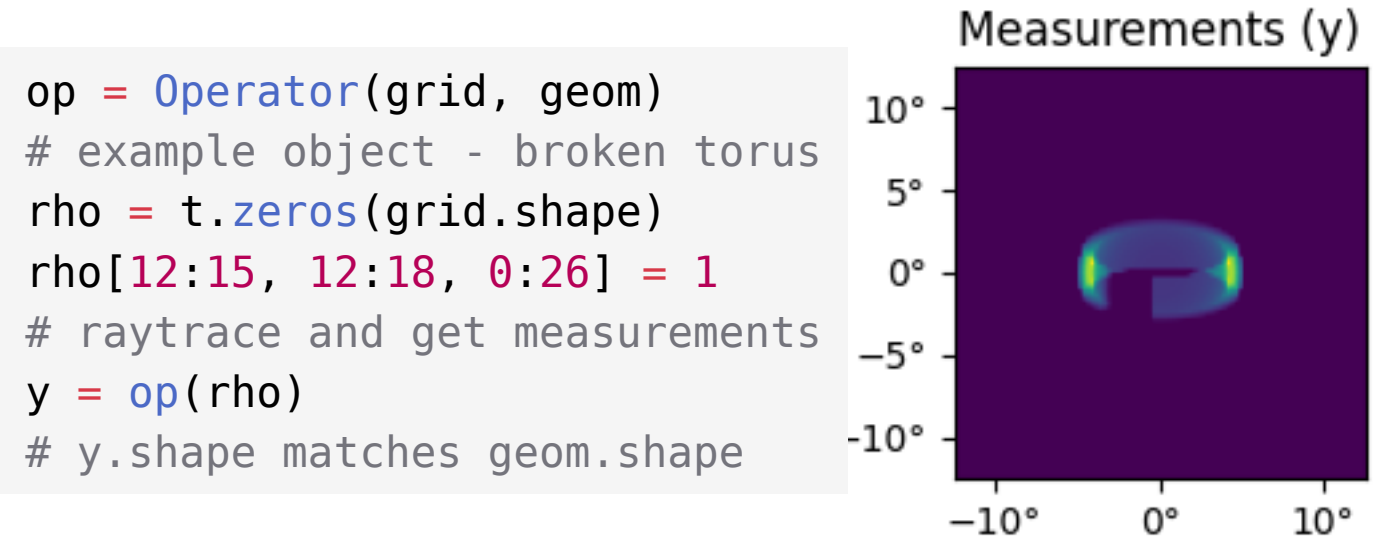

```
op = Operator(grid, geom)
# example object - broken torus
rho = t.zeros(grid.shape)
rho[12:15, 12:18, 0:26] = 1
# raytrace and get measurements
y = op(rho)
# y.shape matches geom.shape
```

Listing 6: Operator constructed from previously defined view geometry and grid with associated measurements

The region index arrays ($r$, $e$, $a$) and length arrays ($\Delta l$) used in Equation 5 are accessible in an `Operator` in the `.regs` and `.lens` attributes, useful when implementing more complicated emission models (see Section 9.3).

### 4.5. Reconstruction Examples

The primary purpose of this library is as a building block for implementing experimental reconstruction algorithms. Listing 7 shows a simple example of a reconstruction algorithm that makes use of gradient descent to reconstruct the example object in the previous section by solving the minimization

$$\hat{\boldsymbol{\rho}} = \arg\min_{\boldsymbol{\rho}} \|\boldsymbol{y} - F\boldsymbol{\rho}\|_2^2 \tag{6}$$

```
# choose initial guess for object
rho_hat = t.zeros(
  grid.shape, device='cuda',
  requires_grad=True
)
# choose optimizer vars and method
optim = t.optim.Adam(
  [rho_hat], lr=1e-3,
  weight_decay=1e2
)
# iterative reconstruction loop.
# minimize L2 loss w.r.t. rho_hat
for i in range(500):
  loss = t.sum((y - op(rho_hat))**2)
  loss.backward()
  optim.step()

y_hat = op(rho_hat)
```

Listing 7: GPU-enabled reconstruction algorithm using off the shelf optimizer from PyTorch and a squared error loss function

Tomographic inverse problems often involve a parametric object model $M$ with low-dimensional set of parameters $\boldsymbol{c}$ and regularization $\mathcal{R}$ of the form

$$\begin{aligned} \hat{\boldsymbol{c}} &= \arg\min_{\boldsymbol{c}} \|\boldsymbol{y} - FM(\boldsymbol{c})\|_2^2 + \mathcal{R}(\boldsymbol{c}) \\ \hat{\boldsymbol{\rho}} &= M(\hat{\boldsymbol{c}}) \end{aligned} \tag{7}$$

which is shown in Listing 8. We emphasize that the model and regularizer can be substituted with any PyTorch-differentiable function with no changes required to the optimization.

```
def model(c):
  # map from low dimensional latent space to R³.
  # use PyTorch-differentiable functions here
  ...
  return rho
def regularizer(c):
  # regularization to enforce behavior on c
  # use PyTorch-differentiable functions here
  ...
  return cost
# choose initial guess for latent variables
c_hat = t.zeros(...)
# choose optimizer vars and method
optim = t.optim.Adam([c_hat], ...)
# iterative reconstruction loop.
# minimize L2 loss w.r.t. c_hat
for i in range(500):
  loss = t.sum((y - op(model(c_hat)))**2)
  loss += regularizer(c_hat)
  ...

rho_hat = model(c_hat)
```

Listing 8: Pseudocode for reconstructing an object that lies in some transform domain defined by `model`

Finally, we remark that the differentiability of the operator allows it to be exploited in machine learning models, such as *physics-informed neural networks* (PINNs), where the operator forms layers of the network itself [31].

### 4.6. Assumptions and Limitations

Unlike most medical imaging paradigms, the detector is located at the vertex of `ConeRectGeom`. We also assume that the detector lies outside of the grid, that rays extend to infinity, and that the object is zero outside of the grid.

When combining view geometries into a collection, the geometries must have the same shape so that the corresponding stack of measurements returned during raytracing is a rectangular array. In the case of a dynamic object, the grid definition must remain constant.

## 5. Raytracer Results

### 5.1. Validation Tests

The raytracer's correctness has been tested by comparison against analytic results for objects where closed-form expressions of column density are available. In the unit sphere test shown in Figure 11, the raytracer computed measurements are within ~0.00001% of the true value, approaching machine epsilon for float32.

Additionally, we have constructed test cases for a variety of LOS/boundary conditions (see Figure 10) which are critical for eliminating bugs in numerically degenerate cases (e.g. a LOS parallel to an azimuthal boundary plane). For each of these cases, we evaluate every step of the raytracer algorithm described in Section 3 to ensure consistency with results computed by hand.

These tests are integrated into the raytracer package's automated testing framework for evaluation when new software releases are made.

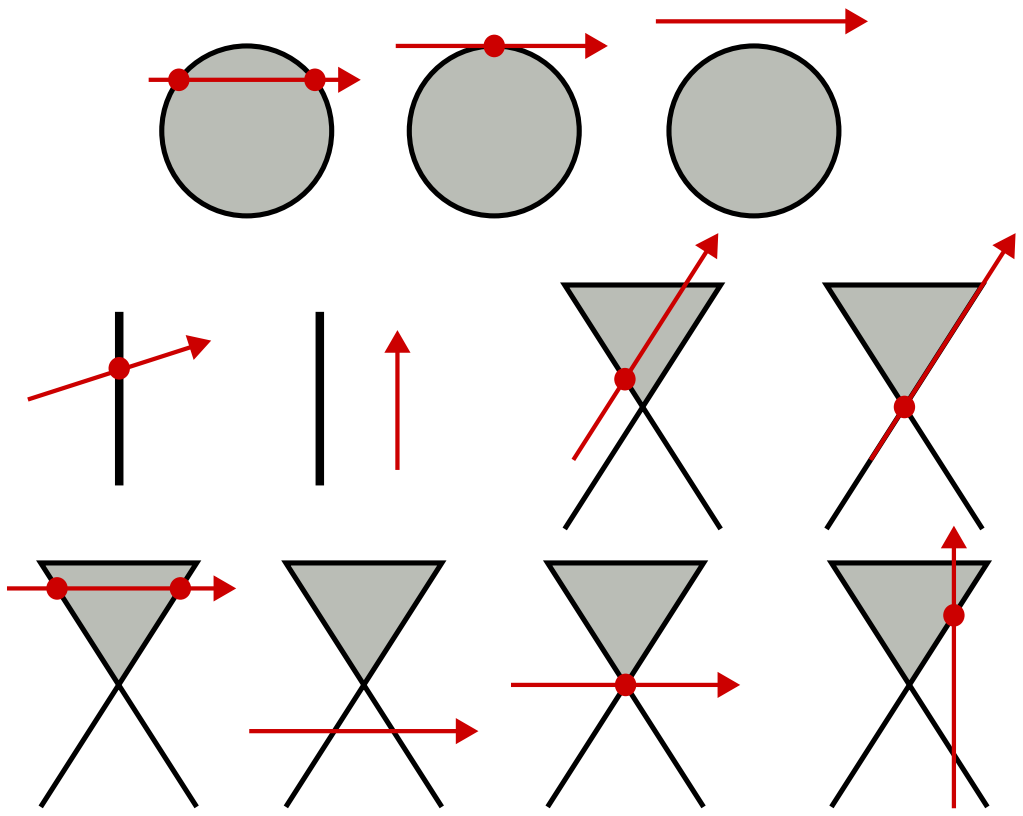

Figure 10: Extensive testing of a combination of various LOS with radial/elevational/azimuthal boundaries. For each case, we compare the output of each step of the raytracer algorithm against analytical results for correctness.

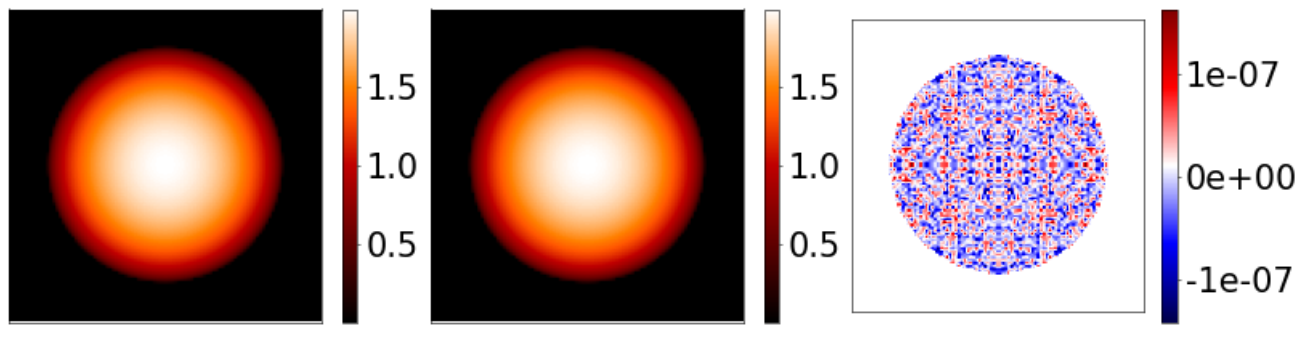

(a) Raytraced (b) Analytic (c) Relative Error

Figure 11: Validation test of raytracer against analytic result for a unit sphere on a (50, 50, 50) grid.

### 5.2. Operator Memory and Timing Benchmarking

Since precomputed arrays are held in memory, it is important to know the peak memory usage of the algorithm to determine whether a particular raytracer forward operator will fit in memory. If running the raytracer on a GPU, memory is even more limited, with most consumer graphics cards constrained to 12GB or 24GB of VRAM as of 2024.

The peak memory required for steps 1-4 can be computed in gigabytes approximately with Equation 8.

$$\text{Peak GB} \approx 6N_{\text{rays}}\,(2N_{\text{r}} + 2N_{\text{e}} + N_{\text{a}})(8)/10^9 \qquad (8)$$

A breakdown of the terms of this expression is below:

- $2N_{\text{r}} + 2N_{\text{e}} + N_{\text{a}}$ - array length (maximum number of voxels intersecting a ray)
- $6N_{\text{rays}}$ - 6 arrays for every ray:
  - voxel indices (rad., elev., azi.)
  - voxel intersection lengths
  - voxel values
  - temporary array for intermediate computation
- 8 - element size of int64 and float64 arrays, in bytes
- $10^9$ - conversion from bytes to gigabytes

Table 3 gives an overview of peak memory use and compute time for several configurations of grid and view geometry shapes. We conducted these tests on Ubuntu 22.04 with AMD Threadripper 5995WX (128GB RAM) and Nvidia RTX 4070 (12GB VRAM) with CUDA 12.4 and PyTorch 2.2.2. Note that additional memory will be consumed by autograd and the PyTorch runtime and may be platform-dependent.

| Num. Rays & Grid Shape | Precompute Time (steps 1-3) | | Raytrace Time (step 4) | | Peak Memory |
|---|---|---|---|---|---|
| | CPU | GPU | CPU | GPU | |
| 1x256x256 (100, 100, 100) | 0.88s | 0.28s | 42ms | 1.5ms | 1.5GB |
| 1x512x512 (50, 50, 50) | 1.5s | 0.17s | 82ms | 3.1ms | 2.9GB |
| 30x64x64 (50, 50, 50) | 0.75s | 0.093s | 37ms | 1.4ms | 1.5GB |

Table 3: Resource usage for various grid and view geometry shapes.

## 6. Tomographic Retrieval of Exospheric Hydrogen

The development of TomoSphero was motivated by the Carruthers Geocorona Observatory, a spacecraft studying the Earth's exosphere that launched in September 2025. An *exosphere* is a region of a planetary atmosphere where density is low enough that particles rarely collide. Earth's exosphere has long been known to primarily be constituted of atomic hydrogen (H), which is detectable by resonant scattering of UV Lyman-α photons from the Sun. This region is of particular interest to physicists because it serves as the pathway for permanent hydrogen escape, important for understanding long-term evolution and historical loss of water from the Earth [32]. Additionally, the presence of an exosphere serves as an indicator of the presence of water on other worlds. As the lower atmosphere is strongly UV-absorbing, measurements must originate from space-based instruments. Carruthers will fill a measurement gap in exospheric science by making continuous wide-field Lyman-α observations from L1 Lagrange point at a distance of approximately 235 Re (Earth radii) over long timescales, as in Figure 12.

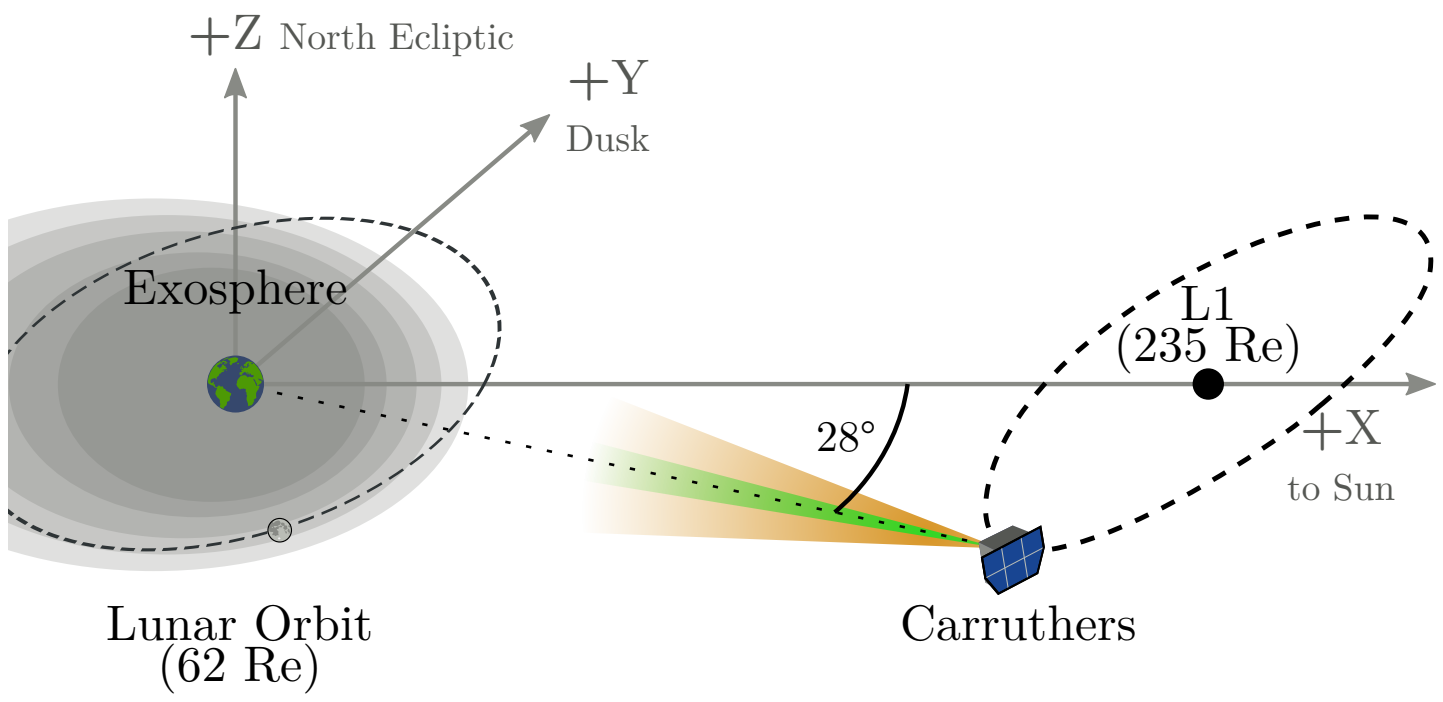

Figure 12: Carruthers 6 month orbit around L1

Equation 9 describes the emission model which relates exospheric H density to Lyman-α photon radiance (phot/s/cm²/sr)

$$y = \frac{g}{4\pi}\phi\left(\vec{n}\right)\int_{l=0}^{\infty}\boldsymbol{\rho}\left(\vec{x}+\vec{n}l\right)\,\mathrm{d}l + \varepsilon \tag{9}$$

where $g$ is solar g-factor (phot/s/atom), $\phi$ is scattering phase function (unitless), $\boldsymbol{\rho}$ is density (atom/cm³), and $\varepsilon$ is measurement noise.

Solar g-factor and scattering phase function together describe the directional rate of emission of photons from an atom, assumed to be known.

We utilize a parametric model for exospheric density in which elevational and azimuthal variations of each radial shell of the density distribution are described by a truncated spherical harmonic series shown in Figure 13.

$$\hat{\boldsymbol{c}} = \arg\min_{\boldsymbol{c}}\|\boldsymbol{y} - FM(\boldsymbol{c})\|_2^2 + \lambda\,\|D_r\boldsymbol{c}\|_2^2$$
$$M(\boldsymbol{c}) \coloneqq \sum_{l=0}^{L}\sum_{m=-l}^{l} c_{rlm}H_{lm}(e,a) \tag{10}$$

where $H_{lm}$ is a spherical harmonic function[2] of degree $l$ and order $m$ defined over elevation and azimuth $(e, a)$, $L$ is the spherical harmonic truncation order, $c_{rlm}$ is a scaling coefficient assigned to each harmonic function on each shell, and $D_r$ is a finite difference operator in the radial direction.

Spherical harmonics are an analog to the Fourier basis on a spherical domain. Truncating to order $L$ limits high-frequency spatial content in the elevational and azimuthal directions, illustrated in Figure 13 up to order $L = 3$. This basis also serves to constrain the low-altitude, noon/midnight region of the exosphere where visibility to Carruthers is blocked by the Earth and optically-thick atmosphere. The second term involving $D_r$ serves as a Tikhonov regularizer to enforce smooth radial decay, controlled by parameter $\lambda$.

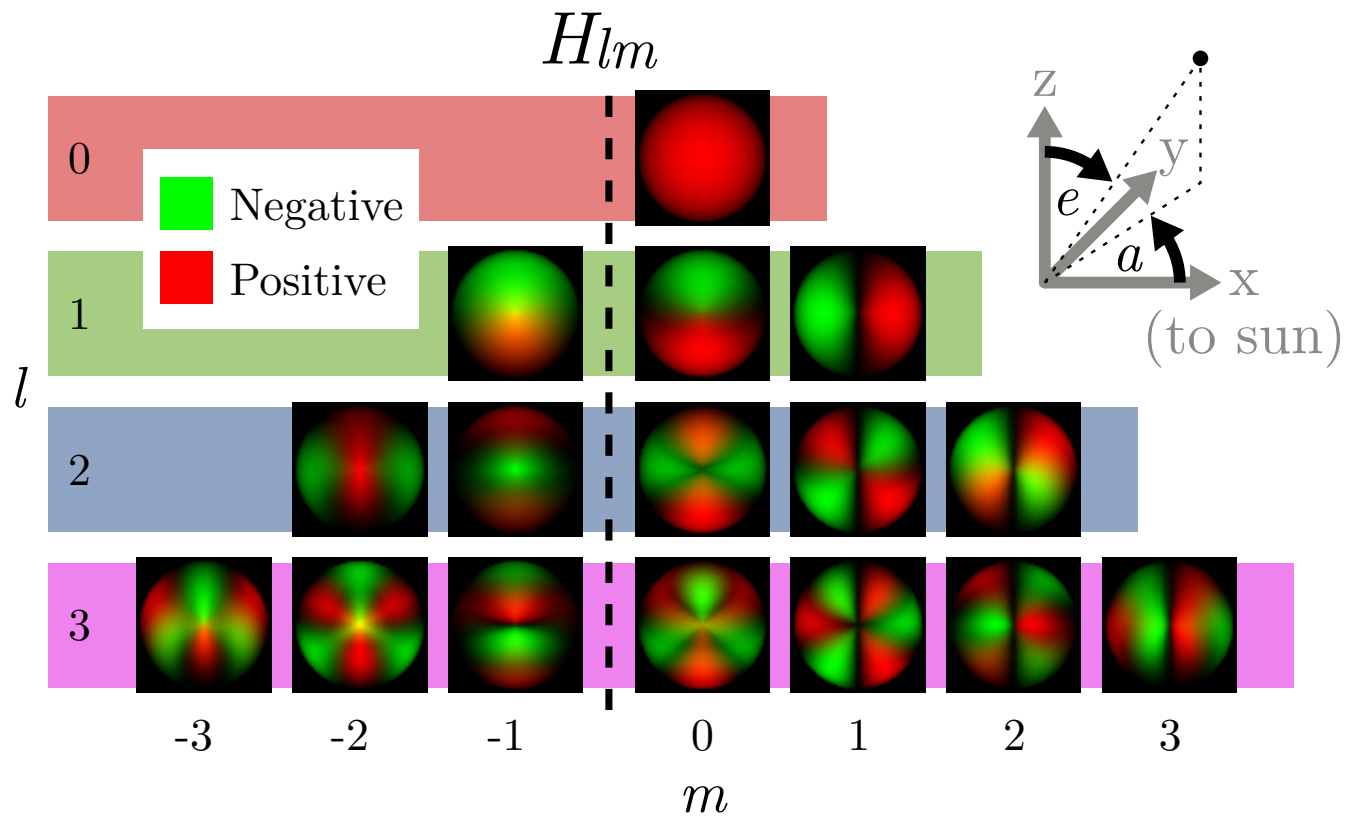

Figure 13: Spherical harmonic functions $H_{lm}$ used by parametric reconstruction model

We used 2015 data during solar max conditions from [33] as a ground truth to test retrieval accuracy of Equation 10 starting at the lower boundary of the optically thin exosphere at 3 Re and extending to 25 Re. We simulated 10 noisy radiance measurements according to Equation 9, spaced uniformly over a 1 month observation window during Spring 2026 according to the projected Carruthers ephemeris. The Carruthers cameras produce rectilinear images[3] which we polar-bin to 100 radial and 50 azimuthal polar pixels with an equivalent FOV. Measurements taken with real cameras are corrupted by many extraneous sources, such as readout noise, particle radiation, shot noise, etc. For Carruthers, these noise sources are well-approximated by 45 dB additive white Gaussian noise in our polar binning scheme. Figure 14 shows the percent error of this 3D retrieval, plotted for slices along the Cartesian planes. The green contour shows the region where density exceeds 25 atoms/cm³, where NASA mission requirements dictate retrieval errors should be less than 50%.

[2] Analogous to Scipy's `sph_harm_y` function

[3] Narrow-field imager: 1024×1024 3.6° FOV, Wide-field imager: 512×512 18° FOV

We note that retrieval errors peak at 19% in the noon/midnight region as expected due to lack of measurement constraints. Retrieval errors are also large near the 25 Re boundary where low H density results in reduced measurement SNR.

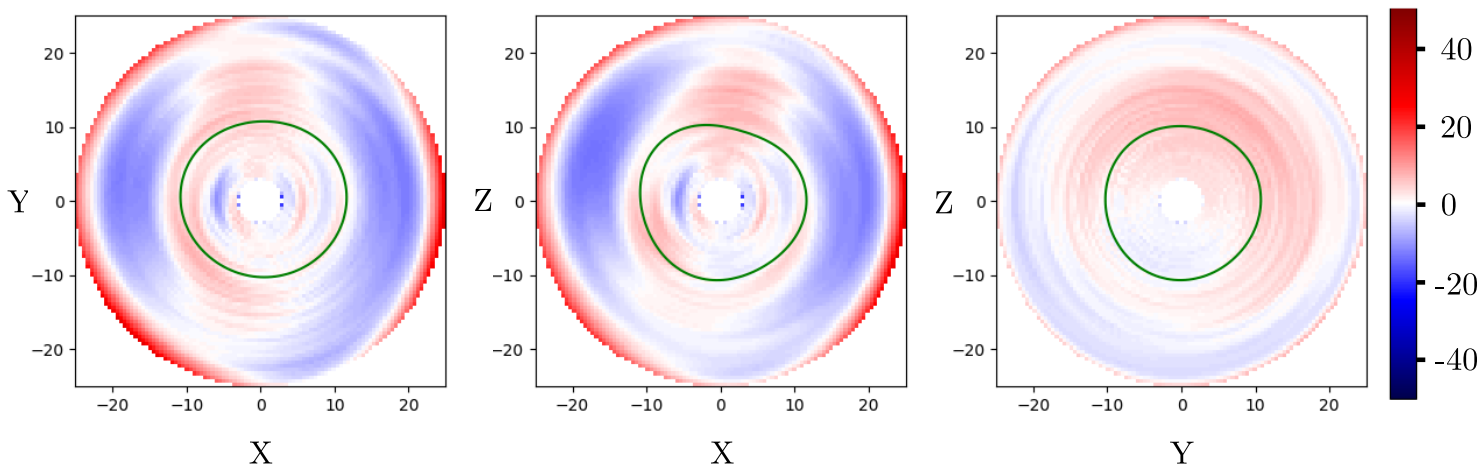

Figure 14: Percent error along Cartesian slices (GSE coordinates) through retrieval using minimization in Equation 10. Green contour demarks 25 atoms/cm³ density boundary

# 7. Conclusion and Future Work

In this work we presented TomoSphero, a GPU-enabled, automatically-differentiable tomographic projector over spherical grids designed for use in reconstruction algorithms. TomoSphero provides visualization tools for validating view geometry correctness and enables rapid development of tomography algorithms in Python that approach the speed of natively-written code, as shown in the benchmarks. We demonstrated the flexibility of automatically-differentiable iterative retrievals via an example of retrieving exospheric H density from simulated noisy radiance measurements, important for the upcoming Carruthers Geocorona Observatory mission.

Future work on TomoSphero may include out-of-core operation to support tomography problems that are larger than GPU memory.

# 8. Acknowledgements

We would like to thank Gonzalo Cucho-Padin for initial feedback on the library API and correction terms. This work was sponsored with the support of NASA grant 17-HPSMO18-1-0020. TomoSphero depends exclusively on PyTorch [16] for numerical computation and Matplotlib [34] for visualization.

# 9. Appendix

## 9.1. Nomenclature and Symbols

- **Grid Boundaries** - spheres, cones, and planes which define the bounds between voxels in a regular spherical grid
- **Ray** - line along which an object will be integrated, a.k.a. "line of sight" (LOS)
- **Grid Region** - area between two adjacent boundaries of the same type
- **View Geometry** - set of lines of sight corresponding to a sensor taking a measurement at a specific location
- **Object** - 3D array of shape $(N_r, N_e, N_a)$ defined on spherical grid holding object density to be integrated
- $\vec{x}$ - starting point of a ray, Cartesian
- $\vec{n}$ - direction of a ray, Cartesian
- $l$ - distance along a ray from the ray starting point
- $\vec{p}$ - intersection point of ray and a boundary, Cartesian
- $\boldsymbol{\rho}$ - 3D object being integrated, continuous or discrete
- $\rho$ - 1D array of object values along a ray
- $\Delta l$ - 1D array of voxel intersection lengths along a ray
- $N$ - number of ray/voxel intersections
- $[r, e, a]$ - indices of a voxel
- $(r, e, a)$ - coordinates of a voxel
- $F$ - forward tomographic operator
- $M$ - parametric object model
- $\boldsymbol{c}$ - low-dimensional model parameters

## 9.2. Boundary Crossing Direction

This section describes how to compute the boundary crossing direction across the three types of grid boundaries as described in Section 3.2. All vectors are in Cartesian coordinates and we assume the principal axis of the spherical grid to be aligned with the Cartesian $+Z$ axis.

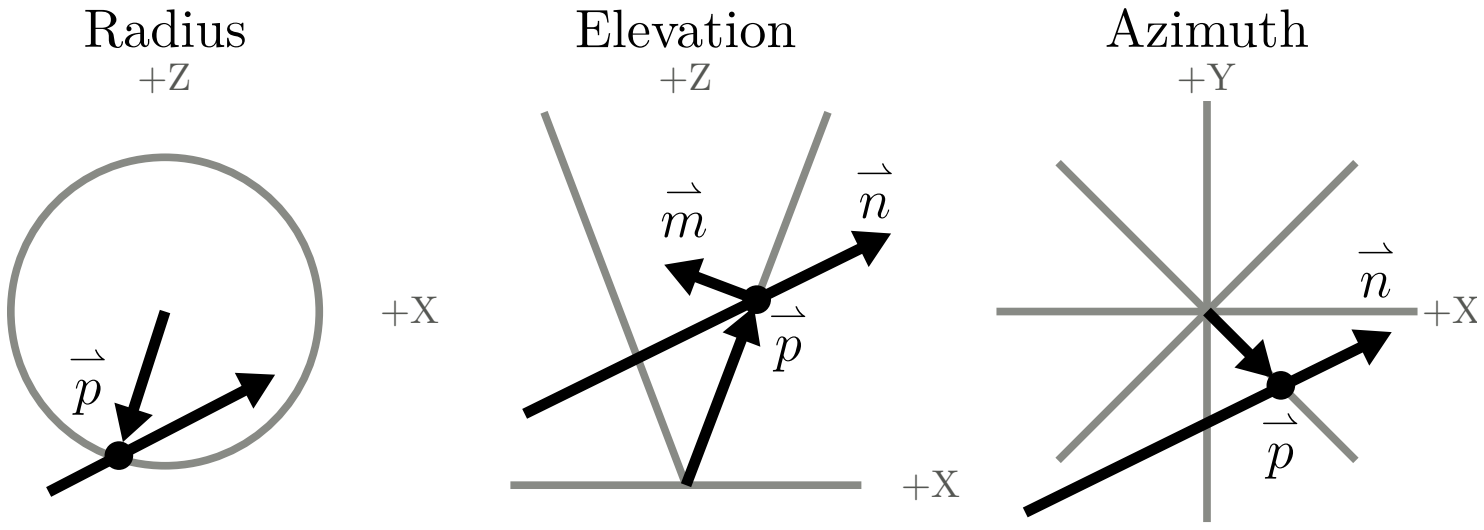

Figure 15: Three types of boundary crossings

Let $\vec{p}$ be the boundary crossing point and $\vec{n}$ be the direction of the ray, as illustrated in Figure 15. To compute the boundary crossing direction over a radial boundary, check for the condition

$$\text{isnegative} = \left(\vec{p} \cdot \vec{n}\right) < 0 \tag{11}$$

For an elevational boundary, check the condition

$$\begin{gathered} \vec{m} = \vec{p} \times \left(-p_y, p_x, 0\right) \\ \text{isnegative} = \left(\vec{m} \cdot \vec{n}\right) > 0 \end{gathered} \tag{12}$$

where $p_x$ and $p_y$ are the $x$ and $y$ components of vector $\vec{p}$, and $\times$ is cross product.

For an azimuthal boundary, check the condition

$$\begin{gathered} \vec{m} = \vec{p} \times \vec{n} \\ \text{isnegative} = m_z < 0 \end{gathered} \tag{13}$$

where $m_z$ is the $z$ component of vector $\vec{m}$.

### 9.3. More Advanced Physics

Zoennchen et al. [33] introduce two first-order correction terms to improve the accuracy of the optically thin approximation described in Section 2.1 by modeling loss of photons as they travel from the Sun to their scattering point and then to the spacecraft, as shown in Figure 16 (a). This section describes how these correction terms may be implemented within the TomoSphero framework.

The first term considers optical depth $\boldsymbol{\tau}_1$ which accounts for self-absorption of solar Lyman-α photons as they travel through the exosphere from the dayside to the nightside along path $S_1$. This results in a reduction of Lyman-α flux by a factor $e^{-\boldsymbol{\tau}_1}$ at scattering point $\vec{s}$.

$$\boldsymbol{\tau}_1\left(\vec{s}\right) = \int_{S_1} (\boldsymbol{\sigma} \cdot \boldsymbol{\rho})\left(\vec{s_1}\right) \mathrm{d}s_1 \tag{14}$$

The second term addresses loss of photons due to secondary scattering that occurs between point $\vec{s}$ and the spacecraft along path $S_2$, related to optical depth $\boldsymbol{\tau}_2$ and resulting in a factor $e^{-\boldsymbol{\tau}_2}$ reduction in photon flux.

$$\boldsymbol{\tau}_2\left(\vec{s}\right) = \int_{S_2} (\boldsymbol{\sigma} \cdot \boldsymbol{\rho})\left(\vec{s_2}\right) \mathrm{d}s_2 \tag{15}$$

Optical depth $\boldsymbol{\tau}_1$ and $\boldsymbol{\tau}_2$ depend on cross section $\boldsymbol{\sigma}$ which is a function of local exospheric temperature, calculated according to Equation 2 in [33].

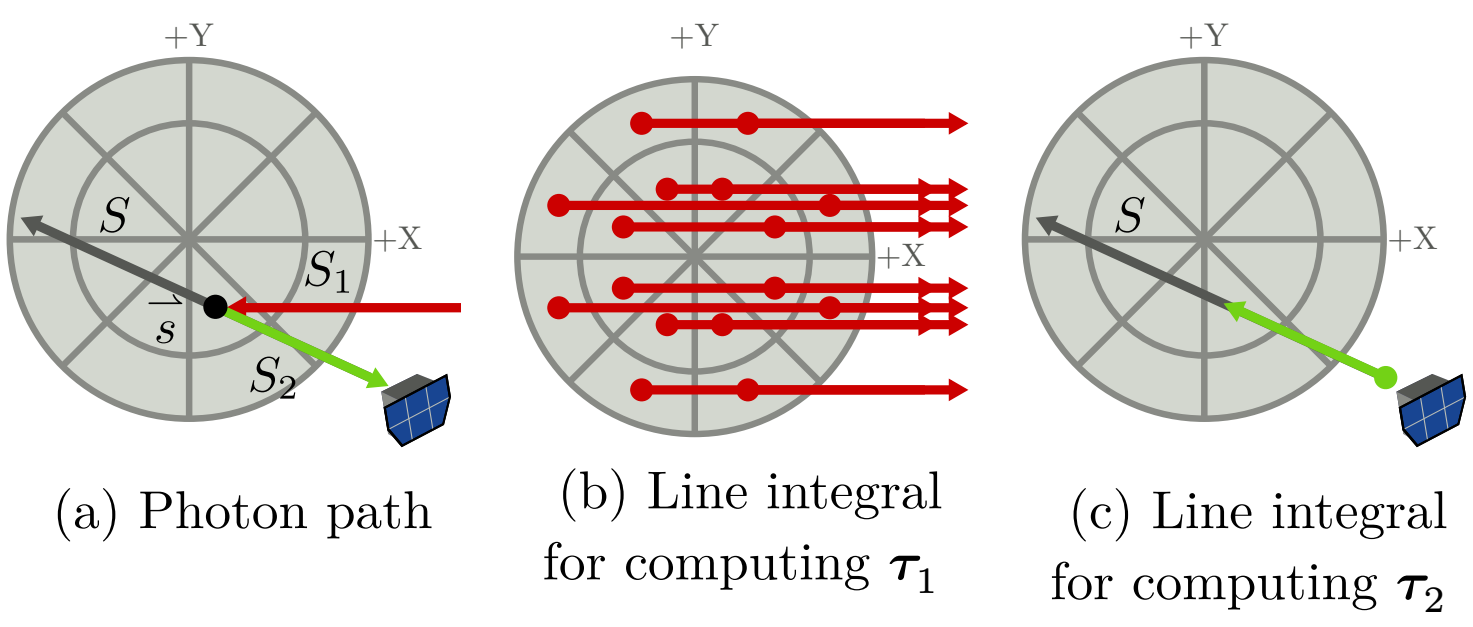

(a) Photon path (b) Line integral for computing $\boldsymbol{\tau}_1$ (c) Line integral for computing $\boldsymbol{\tau}_2$

Figure 16: Correction terms for improving optically-thin approximation

The corrected radiance is

$$y = \frac{g}{4\pi}\phi\left(\vec{n}\right)\int_S e^{-\boldsymbol{\tau}_1\left(\vec{s}\right)} \cdot e^{-\boldsymbol{\tau}_2\left(\vec{s}\right)} \cdot \boldsymbol{\rho}\left(\vec{s}\right) \mathrm{d}s + \varepsilon \tag{16}$$

where $S$ is the LOS parameterized by $\vec{x} + \vec{n}l$ for $l \in [0, \infty]$.

As before, this integral can be discretized as an inner product

$$\begin{aligned} y = F\boldsymbol{\rho} &= \sum_{i=0}^{N-1} (e^{-\boldsymbol{\tau}_1} \odot e^{-\boldsymbol{\tau}_2} \odot \boldsymbol{\rho})[r_i, e_i, a_i] \cdot \Delta l_i \\ &= \langle (e^{-\boldsymbol{\tau}_1} \odot e^{-\boldsymbol{\tau}_2} \odot \boldsymbol{\rho})[r, e, a], \Delta l\rangle \\ &= \langle (e^{-\boldsymbol{\tau}_1} \odot \boldsymbol{\rho})[r, e, a] \odot e^{-\tau_2}, \Delta l\rangle \\ \boldsymbol{\tau}_1 &= F_1(\boldsymbol{\sigma} \odot \boldsymbol{\rho}) \\ \tau_2 &= \mathrm{cumsum}((\boldsymbol{\sigma} \odot \boldsymbol{\rho})[r, e, a] \odot \Delta l) \end{aligned} \tag{17}$$

where $F_1$ is a custom raytracer operator (see Listing 4) with each LOS originating at a voxel centroid and propagating sunwards as in Figure 16 (b), and cumsum is a cumulative summation of a 1D array

$$\mathrm{cumsum}(a)_i = \sum_{j=0}^{i} a_j \tag{18}$$

We note that operator $F_1$ is an approximation of an integral over $S_1$ for each point $\vec{s}$, as voxel centroids do not lie directly on the path $S$.

We obtain $\tau_2$ from $\boldsymbol{\tau}_2$ by only considering points along LOS $S$ to be able to factor the $e^{-\tau_2}$ term out of the product with $\boldsymbol{\rho}$. We also take advantage of the fact that $S_2$ lies along the $S$ associated with $F$ (Figure 16 (c)) to be able to reuse index arrays $r$, $e$, $a$ and length array $\Delta l$.

The sizes of arrays used in this derivation are

- $r, e, a, \Delta l, \tau_2 \in \mathbb{R}^N$
- $\boldsymbol{\rho}, \boldsymbol{\tau}_1, \boldsymbol{\tau}_2, \boldsymbol{\sigma} \in \mathbb{R}^{N_r \times N_e \times N_a}$

# Bibliography

[1] M. Butala, R. Hewett, R. Frazin, and F. Kamalabadi, "Dynamic three-dimensional tomography of the solar corona," *Solar Physics*, vol. 262, pp. 495–509, 2010.

[2] B. Jackson *et al.*, "Three-dimensional reconstruction of heliospheric structure using iterative tomography: A review," *Journal of Atmospheric and Solar-Terrestrial Physics*, vol. 73, no. 10, pp. 1214–1227, 2011, doi: https://doi.org/10.1016/j.jastp.2010.10.007.

[3] D. G. Lloveras *et al.*, "Three-dimensional structure of the corona during WHPI campaign rotations CR-2219 and CR-2223," *Journal of Geophysical Research: Space Physics*, vol. 127, no. 6, p. e2022JA030406, 2022.

[4] M. Kramar, V. Airapetian, Z. Mikić, and J. Davila, "3D coronal density reconstruction and retrieving the magnetic field structure during solar minimum," *Solar Physics*, vol. 289, no. 8, pp. 2927–2944, 2014.

[5] H. Morgan and A. C. Cook, "The width, density, and outflow of solar coronal streamers," *The Astrophysical Journal*, vol. 893, no. 1, p. 57, 2020.

[6] M. J. Willemink and P. B. Noël, "The evolution of image reconstruction for CT—from filtered back projection to artificial intelligence," *European radiology*, vol. 29, pp. 2185–2195, 2019.

[7] H. A. Gabbar, A. Chahid, M. J. A. Khan, O. G. Adegboro, and M. I. Samson, "CTIMS: Automated Defect Detection Framework Using Computed Tomography," *Applied Sciences*, vol. 12, no. 4, 2022, doi: 10.3390/app12042175.

[8] H. Nass, J. Zoennchen, G. Lay, and H. Fahr, "The TWINS-LAD mission: Observations of terrestrial Lyman-α fluxes," *Astrophysics and Space Sciences Transactions*, vol. 2, no. 1, pp. 27–31, 2006.

[9] F. S. Prol, M. M. Hoque, and A. A. Ferreira, "Plasmasphere and topside ionosphere reconstruction using

METOP satellite data during geomagnetic storms," *Journal of Space Weather and Space Climate*, vol. 11, p. 5, 2021.

[10] S. Pryse, L. Kersley, D. Rice, C. Russell, and I. Walker, "Tomographic imaging of the ionospheric mid-latitude trough," in *Annales Geophysicae*, 1993, pp. 144–149.

[11] E. Y. Sidky and X. Pan, "Image reconstruction in circular cone-beam computed tomography by constrained, total-variation minimization," *Physics in Medicine & Biology*, vol. 53, no. 17, p. 4777, 2008.

[12] J. Austen, S. Franke, C. Liu, and K. Yeh, "Application of computerized tomography techniques to ionospheric research," in *International beacon satellite symposium on radio beacon contribution to the study of ionization and dynamics of the ionosphere and to corrections to geodesy and technical workshop*, 1986, pp. 25–35.

[13] J. A. Scales, "Tomographic inversion via the conjugate gradient method," *Geophysics*, vol. 52, no. 2, pp. 179–185, 1987.

[14] S. V. Venkatakrishnan, C. A. Bouman, and B. Wohlberg, "Plug-and-Play priors for model based reconstruction," in *2013 IEEE Global Conference on Signal and Information Processing*, 2013, pp. 945–948. doi: 10.1109/GlobalSIP.2013.6737048.

[15] C. Thibaudeau, J.-D. Leroux, R. Fontaine, and R. Lecomte, "Fully 3D iterative CT reconstruction using polar coordinates," *Medical physics*, vol. 40, no. 11, p. 111904, 2013.

[16] A. Paszke *et al.*, "PyTorch: An Imperative Style, High-Performance Deep Learning Library," *Advances in Neural Information Processing Systems 32*. Curran Associates, Inc., pp. 8024–8035, 2019. [Online]. Available: http://papers.neurips.cc/paper/9015-pytorch-an-imperative-style-high-performance-deep-learning-library.pdf

[17] J. Bradbury *et al.*, "JAX: composable transformations of Python+NumPy programs." [Online]. Available: http://github.com/google/jax

[18] A. Biguri, M. Dosanjh, S. Hancock, and M. Soleimani, "TIGRE: a MATLAB-GPU toolbox for CBCT image reconstruction," *Biomedical Physics & Engineering Express*, vol. 2, no. 5, p. 55010, Sept. 2016, doi: 10.1088/2057-1976/2/5/055010.

[19] H. Kim and K. Champley, "Differentiable Forward Projector for X-ray Computed Tomography." [Online]. Available: https://arxiv.org/abs/2307.05801

[20] W. van Aarle *et al.*, "Fast and flexible X-ray tomography using the ASTRA toolbox," *Opt. Express*, vol. 24, no. 22, pp. 25129–25147, Oct. 2016, doi: 10.1364/OE.24.025129.

[21] W. van Aarle *et al.*, "The ASTRA Toolbox: A platform for advanced algorithm development in electron tomography," *Ultramicroscopy*, vol. 157, pp. 35–47, 2015, doi: https://doi.org/10.1016/j.ultramic.2015.05.002.

[22] W. Palenstijn, K. Batenburg, and J. Sijbers, "Performance improvements for iterative electron tomography reconstruction using graphics processing units (GPUs)," *Journal of Structural Biology*, vol. 176, no. 2, pp. 250–253, 2011, doi: https://doi.org/10.1016/j.jsb.2011.07.017.

[23] C. A. Bouman and G. T. Buzzard, "MBIRJAX: High-performance tomographic reconstruction." 2024.

[24] D. Kazantsev and N. Wadeson, "TOmographic MOdel-BAsed Reconstruction (ToMoBAR) software for high resolution synchrotron X-ray tomography," in *CT Meeting*, 2020.

[25] J. S. Jørgensen *et al.*, "Core Imaging Library-Part I: a versatile Python framework for tomographic imaging," *Philosophical Transactions of the Royal Society A*, vol. 379, no. 2204, p. 20200192, 2021.

[26] A. Hendriksen *et al.*, "Tomosipo: Fast, Flexible, and Convenient 3D Tomography for Complex Scanning Geometries in Python," *Optics Express*, Oct. 2021, doi: 10.1364/oe.439909.

[27] G. Cucho-Padin and L. Waldrop, "Tomographic estimation of exospheric hydrogen density distributions," *Journal of Geophysical Research: Space Physics*, vol. 123, no. 6, pp. 5119–5139, 2018.

[28] J. Guertault, "Intersection of a ray and a cone." Accessed: Feb. 01, 2024. [Online]. Available: http://lousodrome.net/blog/light/2017/01/03/intersection-of-a-ray-and-a-cone/

[29] D. Eberly, "Intersection of a Line and a Cone." Accessed: Feb. 01, 2024. [Online]. Available: https://www.geometrictools.com/Documentation/IntersectionLineCone.pdf

[30] K. Halliday, "Ray-Sphere Intersection with Simple Math." Accessed: Feb. 01, 2024. [Online]. Available: https://kylehalladay.com/blog/tutorial/math/2013/12/24/Ray-Sphere-Intersection.html

[31] M. Raissi, P. Perdikaris, and G. E. Karniadakis, "Physics-informed neural networks: A deep learning framework for solving forward and inverse problems involving nonlinear partial differential equations," *Journal of Computational physics*, vol. 378, pp. 686–707, 2019.

[32] I. I. Baliukin, J.-L. Bertaux, E. Quémerais, V. Izmodenov, and W. Schmidt, "SWAN/SOHO Lyman-α mapping: The hydrogen geocorona extends well beyond the Moon," *Journal of Geophysical Research: Space Physics*, vol. 124, no. 2, pp. 861–885, 2019.

[33] J. H. Zoennchen, G. Cucho-Padin, L. Waldrop, and H. J. Fahr, "Comparison of terrestrial exospheric hydrogen 3D distributions at solar minimum and maximum using

TWINS Lyman-alpha observations," *Frontiers in Astronomy and Space Sciences*, vol. 11, p. 1409744, 2024.

[34] J. D. Hunter, "Matplotlib: A 2D graphics environment," *Computing in Science & Engineering*, vol. 9, no. 3, pp. 90–95, 2007, doi: 10.1109/MCSE.2007.55.